\documentclass[usenatbib]{mn2e}

\usepackage{epsfig}
\usepackage{mathrsfs,aas_macros}

\def\Lya{Ly$\alpha$~}

\def\HI{\hbox{H$\,\rm \scriptstyle I\ $}}
\def\HII{\hbox{H$\,\rm \scriptstyle II\ $}}

\title[The rapid demise of \Lya emitters]{On the rapid demise of
  Lyman-$\alpha$ emitters at redshift \boldmath{$z \ga 7$} due to the
  increasing incidence of optically thick absorption systems}

\author[J.S. Bolton \& M.G. Haehnelt] {James S. Bolton$^{1,2}$ \&
  Martin G. Haehnelt$^{3}$\\$^1$ School of Physics and Astronomy,
  University of Nottingham, University Park, Nottingham, NG7 2RD
  \\$^2$ School of Physics, University of Melbourne, Parkville, VIC
  3010, Australia \\$^3$ Kavli Institute for Cosmology and Institute
  of Astronomy, Madingley Road, Cambridge, CB3 0HA }

\begin{document}

\date{\today}

\maketitle

\label{firstpage}

\begin{abstract}

A variety of independent observational studies have now reported a
significant decline in the fraction of Lyman-break galaxies which
exhibit \Lya emission over the redshift interval $z=6$--$7$.  In
combination with the strong damping wing extending redward of \Lya in
the spectrum of the bright $z=7.085$ quasar ULAS 1120$+$0641, this has
strengthened suggestions that the hydrogen in the intergalactic medium
(IGM) is still substantially neutral at $z\sim 7$.  Current
theoretical models imply \HI fractions as large as $40$--$90$ per cent
may be required to explain these data assuming there is no intrinsic
evolution in the \Lya emitter population.  We propose that such large
neutral fractions are not necessary.  Based on a hydrodynamical
simulation which reproduces the absorption spectra of high-redshift
($z\sim 6-7$) quasars, we demonstrate that the opacity of the
intervening IGM redward of rest-frame \Lya can rise rapidly in average
regions of the Universe simply because of the increasing incidence of
absorption systems which are optically thick to Lyman continuum
photons as the tail-end of reionisation is approached.  Our
simulations suggest these data do not require a large change in the
IGM neutral fraction by several tens of per cent from $z=6$--$7$, but may
instead be indicative of the rapid decrease in the typical mean free
path for ionising photons expected during the final stages of
reionisation.
\end{abstract}


%%%%%%%%%%%%%%%%%%%%%%%%%%%%%%%%%%%%%%%%%%%%%%%%%%%%%%%%%%%%%%%%%%%%%	
%%%%%%%%%%%%%%%%%%%%%%%%%% SECTION 1 %%%%%%%%%%%%%%%%%%%%%%%%%%%%%%%%
%%%%%%%%%%%%%%%%%%%%%%%%%%%%%%%%%%%%%%%%%%%%%%%%%%%%%%%%%%%%%%%%%%%%%

\section{Introduction}
 
The opacity observed blueward of rest frame \Lya in the spectra of
distant quasars rises toward higher redshifts
(\citealt{Fan06b,Becker07}), indicating the fraction of neutral
hydrogen in the intergalactic medium (IGM) is small but increasing
with lookback time.  This observation, coupled with the Thomson
optical depth measured from cosmic microwave background data
(\citealt{Komatsu11}), suggests that an extended epoch of hydrogen
reionisation was ending by $z \sim 6-7$.  Reaching further into the
epoch of reionisation -- when intergalactic hydrogen is still
substantially neutral -- is difficult, but \Lya selected galaxies may
offer one promising route to probing somewhat deeper into this distant
era (\citealt{MiraldaEscudeRees98,Haiman02}).  The increasing \HI
content in the IGM at $z>6$ produces a \Lya damping wing which can
extend redward of a galaxy's \Lya emission line. The damping wing
reduces the visibility of the \Lya emission
(\citealt{MiraldaEscude98}), and the number of \Lya emitting galaxies
(LAEs) observed in flux-limited surveys will thus decrease as the
ambient \HI fraction rises with increasing redshift
(\citealt{HaimanSpaans99}).

The resonant nature of the \Lya transmission unfortunately complicates
this simple picture.  Source clustering (\citealt{Furlanetto06}), dust
(\citealt{HansenOh06,Dayal11}), and resonant \Lya scattering within
the sources themselves -- which is sensitive to both the \HI content
and gas velocities (\citealt{Santos04,Dijkstra10}) -- all impact on
the visibility of \Lya emission. The complex resonant radiative
transfer within the galaxies often leads to substantial line redshifts
of several hundred $\rm km\,s^{-1}$ which can significantly enhance
the visibility of LAEs when the surrounding IGM is still substantially
neutral (\citealt{Dijkstra11,Laursen11,Zheng11,Barnes11,Jeeson12}).

Recently, however, a significant decrease in the transmission of \Lya
photons from high redshift galaxies has been found between $z=6$--$7$,
(\citealt{Stark10,Pentericci11,Hayes11,Ono12,Schenker12,CurtisLake12}).
This result is based on the rapid decline in the fraction of Lyman
break galaxies (LBGs) which exhibit \Lya emission -- a quantity which
is (in principle) less susceptible than the LAE luminosity function
(e.g. \citealt{MalhotraRhoads04,Kashikawa06,Hu10}) to observational
selection effects and intrinsic evolution in the LAE population.
Based on comparisons to existing numerical simulations and
semi-numerical models of reionisation, the reported rapid decrease in
the transmission of \Lya emission at $z\sim 6 -7$ may be interpreted
as a rapid change in the volume-averaged neutral fraction (by several
tens of percent) over a rather short redshift interval
(\citealt{Pentericci11,Schenker12,Ono12}).

Such a rapid change in the neutral fraction is, however, at odds with
the rather low ionising emissivity suggested by the \Lya forest data
at $z=5$--$6$ (\citealt{BoltonHaehnelt07b}) which appears to imply a
much slower evolution of the neutral fraction.  Theoretical
reionisation models matching the \Lya forest data
(e.g. \citealt{Ciardi12,Kuhlen12,Jensen12}) therefore have
considerable difficulty in explaining the rather large neutral
fractions ($\sim 40$--$90$ per cent) that have been inferred from the
recent LAE/LBG observations at $z\sim 7$.  Alternative explanations
which -- either individually or in combination -- might allow for a
more modest change in the IGM neutral fraction include an increase in
the escape fraction of ionising photons, or an increase in the
interstellar dust content of the galaxies toward higher redshift (see
\citealt{ForeroRomero12} for further discussion of these points).

In this paper, we shall argue that one other clue to this puzzle is
provided by the quasar ULAS J1120$+$0641 at $z=7.085$, recently
discovered by the UKIRT Infrared Deep Sky Survey (UKIDSS,
\citealt{Lawrence07}).  The spectrum of ULAS J1120$+$0641 exhibits a
strong \Lya damping wing extending redward of the quasar \Lya emission
line (\citealt{Mortlock11}) -- exactly what is needed to suppress the
generally redshifted (relative to the systemic redshift of the galaxy)
emission of LAEs.  Theoretical models predict that the regions
surrounding rare, bright quasars are amongst the first to reionise
(e.g. \citealt{Furlanetto04b,Iliev06b,TracCen07,Zahn07,McQuinn07}).
If the environment around ULAS J1120$+$0641 is typical of quasar host
galaxies at $z>7$, the significantly weaker ionising radiation field
expected around even the brightest LAEs implies the impact of the red
\Lya damping wing on LAE visibility should be even stronger.

\cite{Bolton11} used high resolution simulations of \Lya absorption to
demonstrate that unless a proximate damped \Lya absorber lies within
$\sim 5\rm \,pMpc$ of the quasar, a volume averaged \HI fraction of at
least $\langle f_{\rm HI}\rangle_{\rm V}=0.1$ is required to reproduce
the damping wing observed around ULAS J1120$+$0641.  In this work we
show that an \HI fraction of around 10 per cent could also
explain the rapid evolution of the LAEs -- obviating the need for \HI
fractions as large as $40$--$90$ per cent.  The difference in the
inferred neutral fraction is due to the inclusion of small-scale
($\sim 20\rm\, pkpc$), optically thick absorptions systems which
remain unresolved in large-scale ($\ga 100 \rm\,cMpc$) reionisation
simulations.  On including these systems the \Lya opacity up to a few
hundred $\rm km\,s^{-1}$ redward of rest frame \Lya rises more rapidly
with \HI fraction than usually predicted.  As we shall demonstrate,
this is because the intervening \Lya opacity of the IGM at the end of
hydrogen reionisation is strongly affected by the increasing size of
these systems, which naturally occurs as the metagalactic
photo-ionisation rate decreases.  A more modest neutral fraction of
$\sim 10$ per cent is much easier to reconcile with the fact that
quasar absorption spectra at $z\simeq 6$ appear to indicate that the
IGM is highly ionised only $180\rm\,Myr$ later
(\citealt{Fan06b,Wyithe08}, but see also \citealt{Mesinger10}).

We begin our analysis in Section 2, where we describe our modelling of
the IGM \Lya opacity and the procedure we adopt for incorporating
self-shielded gas into a hydrodynamical simulation.  In Section 3 the
properties of the optically thick absorbers in the simulations are
discussed in more detail.  We explore the implications of these
absorption systems for the visibility of \Lya emission from galaxies
in Section 4, and in Section 5 we conclude.  We assume the
cosmological parameters $\Omega_{\rm m}=0.26$,
$\Omega_{\Lambda}=0.74$, $\Omega_{\rm b}h^{2}=0.023$, $h=0.72$ and a
helium mass fraction of $Y=0.24$ throughout.  Comoving and proper
distances are denoted by using the prefixes ``c'' and ``p'',
respectively.

\section{Numerical simulations of the IGM}
\subsection{Hydrodynamical model}

\begin{figure*}
\centering
\begin{minipage}{180mm}
\begin{center}
 \psfig{figure=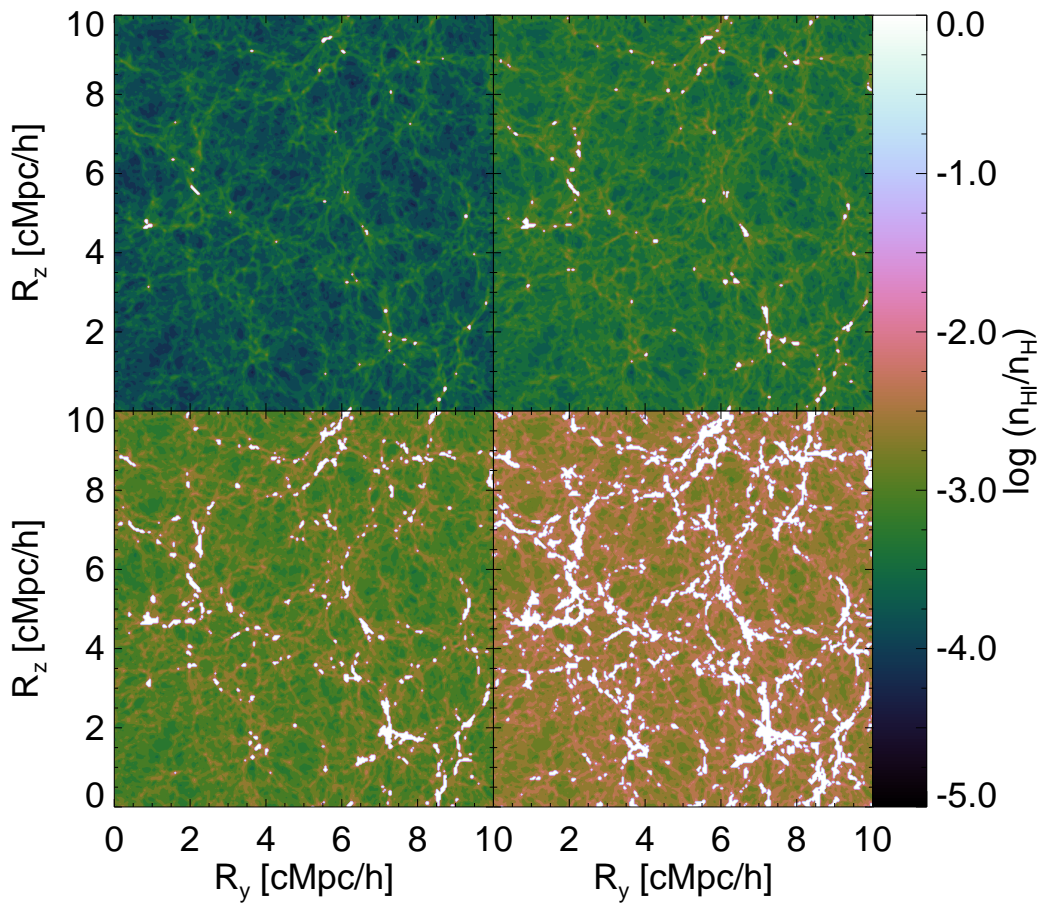,width=0.8\textwidth}
 \vspace{-0.3cm}

 \caption{Slices midway ($R_{\rm x}=5h^{-1}\rm \,cMpc$, with width
   $\Delta R_{\rm x}=39h^{-1}\rm\,ckpc$) through the hydrodynamical
   simulation volume at $z=7$ displaying the spatial distribution of
   the neutral hydrogen fraction, $n_{\rm HI}/n_{\rm H}$, for four
   different background photo-ionisation rates.  A self-shielding
   prescription (see Eq.~\ref{eq:shield}) has been used in all four
   cases.  Fully neutral self-shielded regions, which trace the
   highest density regions in the simulation, are shaded white.
   Anti-clockwise from top left, $\log(\Gamma_{\rm HI}/\rm
   s^{-1})=-12.8,\,-13.6,\,-14.0$ and $-13.2$.}
\label{fig:slices}

\end{center}
\end{minipage}
\end{figure*}

We model the IGM using a high-resolution hydrodynamical simulation
performed with the parallel Tree-SPH code {\tt GADGET-3}, last
described in \cite{Springel05}.  The simulation employs a box size of
$10h^{-1}$ cMpc, a gas particle mass of $9.2\times
10^{4}h^{-1}M_{\odot}$ and a gravitational softening length of
$0.65h^{-1}\rm\,ckpc$.  The gas is reionised instantaneously by a
uniform, optically thin ionising background at $z=9$ based on the
\cite{HaardtMadau01} galaxies and quasars emission model.  Although we
subsequently recompute the ionisation state of the gas, this means the
IGM density field at $z\leq 9$ is already Jeans smoothed by the
increased gas pressure following photo-heating
(e.g. \citealt{Pawlik09}).  This simulation was previously used to
model the \Lya opacity in the near-zone of the high-redshift quasar
ULAS J1120$+$0641 at $z=7.085$ (\citealt{Bolton11}).  In this work we
use an output at $z=7$.

The initial equilibrium ionisation state of the gas is set by a
spatially uniform ionising background.  We do not model the patchy
ionisation state of the IGM by following the radiative transfer of
ionising photons emitted by individual sources.  We will, however,
estimate the effect of a central source on the ionisation state of the
IGM later in our analysis.  A box size of $10h^{-1}$ cMpc is in any
case too small to correctly capture the large scale topology of the
reionisation process.  Instead, the high mass resolution of the
simulation allows us to incorporate the small-scale structures
responsible for optically thick absorption systems.  These absorption
systems are expected to define the edge of ionised bubbles
(\citealt{Crociani11}), regulate the mean free path at overlap
(\citealt{MiraldaEscude00,GnedinFan06,McQuinn11}) and delay the latter
stages of reionisation (\citealt{Ciardi06,AlvarezAbel12}).  These
systems remain unresolved in most large-scale reionisation simulations
(but see \citealt{Choudhury09, Crociani11, AlvarezAbel12} for recent
``semi-numerical'' approaches, and
\citealt{GnedinFan06,Kohler07,McQuinn11,Altay11} for smaller
simulations which resolve self-shielded systems) and are therefore
often not considered when estimating the impact of intergalactic \Lya
absorption on the visibility of LAEs.

\subsection{Modelling optically thick absorption systems}

In order to include optically thick absorbers in our model, following
\cite{Schaye01} we use a simple, physically motivated prescription to
assign a density threshold above which gas remains self-shielded,
which we briefly summarise here (see also
\citealt{MiraldaEscude00,FurlanettoOh05}).  Assuming that the typical
size of an \HI absorber is the Jeans scale and adopting the case-A
recombination coefficient for ionised hydrogen (\citealt{Abel97}), the
overdensity at which an \HI absorber begins to
self-shield may be approximated as:

\begin{equation} \Delta_{\rm ss} = 36\, \Gamma_{-12}^{2/3}T_{4}^{2/15}\left(\frac{\mu}{0.61}\right)^{1/3}\left(\frac{f_{\rm e}}{1.08}\right)^{-2/3}\left(\frac{1+z}{8}\right)^{-3}, \label{eq:shield} \end{equation}

\noindent
where $\Gamma_{-12}=\Gamma_{\rm HI}/10^{-12}\rm\,s^{-1}$ is the
background photo-ionisation rate, $T_{4}=T/10^{4}\rm\,K$ is the gas
temperature, $\mu$ is the mean molecular weight and $f_{\rm e}=n_{\rm
  e}/n_{\rm H}$ is the free electron fraction with respect to
hydrogen.  Absorbers at this density threshold have a characteristic
size

\begin{equation} L_{\rm ss}=8.7{\rm\,pkpc}\,\Gamma_{-12}^{-1/3}T_{4}^{13/30}\left(\frac{\mu}{0.61}\right)^{-2/3}\left(\frac{f_{\rm e}}{1.08}\right)^{1/3}. \label{eq:Jeans} \end{equation}

\noindent 
The fiducial values adopted for $f_{\rm e}$ and $\mu$ correspond to a
plasma of ionised hydrogen and singly ionised helium.  Note that
helium is not expected to be doubly ionised until $z\sim 3$
(e.g. \citealt{FurlanettoOh08b,McQuinn09}).  Comparison to radiative
transfer calculations indicate this simple prescription recovers the
density at which intergalactic gas self-shields to within a factor of
two at $z=6$ (\citealt{McQuinn11}).

We shall consider four different values for the background
photo-ionisation rate in this work: $\log(\Gamma_{\rm HI}/{\rm
  s^{-1}})=-12.8,\,-13.2,\,-13.6$ and $-14.0$.  These are summarised
in Table~\ref{tab:uvb} along with the corresponding volume weighted
\HI fractions.  This range is chosen so that the largest value is
comparable to recent photo-ionisation rate measurements from the \Lya
forest at $z=6$ (\citealt{WyitheBolton11,Calverley11}).  The lowest
value produces a volume averaged \HI fraction of around ten per cent
(with $\Delta_{\rm ss}\sim 2$), and thus approaches the regime where
this self-shielding prescription ceases to provide an adequate
description\footnote{In other words, when the IGM is no longer in the
  ``post-overlap'' stage of reionisation (cf. \citealt{Gnedin00}) and
  neutral gas also resides in underdense regions far from ionising
  sources.  This self-shielding model only applies to what is
  sometimes referred to as the final ``outside-in'' stages of
  reionisation
  (e.g. \citealt{MiraldaEscude00,Choudhury09,Finlator09}).} of the
spatial distribution of \HI in the IGM.

\begin{table}
  \centering
  \caption{The four background photo-ionisation rates used in this
    work.  The resulting volume weighted neutral hydrogen fractions
    (which include the contribution from self-shielded gas) are
    summarised in the second column.}
  \begin{tabular}{c|c}
    \hline
    $\log(\Gamma_{\rm HI}/\rm s^{-1})$ & $\langle f_{\rm HI} \rangle_{\rm V}$  \\
    \hline
    -12.8  & $2.7\times 10^{-3}$  \\
    -13.2  & $9.2 \times 10^{-3}$ \\
    -13.6  & $3.2 \times 10^{-2}$ \\
    -14.0  & $1.1 \times 10^{-1}$ \\
    \hline
  \end{tabular}
\label{tab:uvb}
\end{table}

Figure~\ref{fig:slices} displays the resulting distribution of neutral
hydrogen in a two dimensional slice of the simulation.  The
self-shielding threshold given by Eq.~(\ref{eq:shield}) is computed
self-consistently using the temperatures and densities from the
hydrodynamical simulation.  Neutral hydrogen (shaded white) traces the
densest parts of the cosmic-web, and fills a progressively larger
fraction of the simulation volume as the amplitude of the ionising
background is lowered.

\subsection{Extracting line-of-sight quantities} \label{sec:los}

We use these neutral hydrogen distributions as the starting point for
constructing mock \Lya absorption spectra.  Sight-lines are extracted
parallel to the box boundaries in three directions around the 100 most
massive dark matter haloes in the simulation at $z=7$.  These halo
sight-lines are spliced together with sight-lines drawn randomly from
the periodic volume to form a total of 300 continuous skewers through
the density, peculiar velocity and temperature field of the
hydrodynamical simulation.  The skewers have a total length of
$12.5\rm\, pMpc$, ranging from $8330\rm\,km\,s^{-1}$ to
$-2084\rm\,km\,s^{-1}$ around the halo position.  The \Lya optical
depth along each of the sight-lines is then computed following a
standard procedure, see e.g.  eq. (15) in \cite{BoltonHaehnelt07}.  In
the rest of our analysis, all quantities will be estimated from these
300 sight-lines.

We note, however, that the most massive dark matter halo in the
simulation volume at $z=7$ has a mass $M_{\rm dm}=9.0\times
10^{9}M_{\odot}$.  In comparison, \cite{Ouchi10} infer an average dark
matter host halo mass of $10^{10}$--$10^{11}\rm\,M_{\odot}$ from the
clustering of LAEs with $L_{\rm Ly\alpha}>2.5\times
10^{42}\rm\,erg\,s^{-1}$ at $z=6.6$.  Our simulation is therefore too
small to correctly sample typical LAE host haloes while simultaneously
resolving the small-scale structure of the IGM, so in what follows we
will underestimate the impact of the more extended gas overdensities
and/or stronger inflows expected around larger haloes on the
visibility of the \Lya emitters.  However, both of these effects will
tend to make the impact of \Lya absorption on the visibility of the
\Lya emitters stronger, rather than weaker, serving to strengthen our
main conclusion that modest \HI fractions may explain the observed LAE
fraction at $z=7$.

\section{Properties of the optically thick absorbers}

\begin{figure}
\begin{center}
  \includegraphics[width=0.5\textwidth]{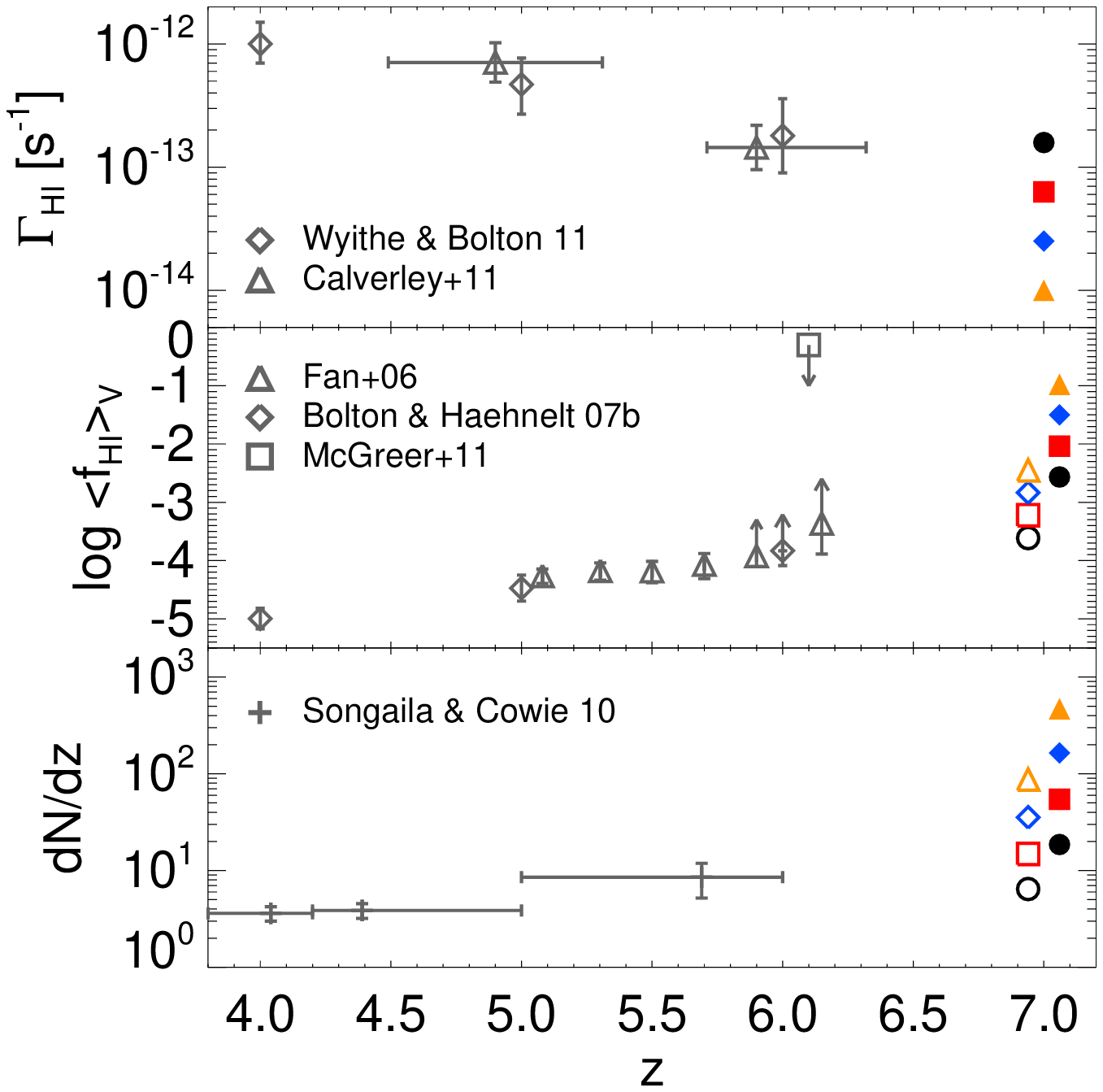}
  \vspace{-0.6cm}
  \caption{{\it Top:} comparison of the four photo-ionisation rates
    assumed in this work (filled symbols at $z=7$) with observational
    constraints at $z\leq 6$ from the \Lya forest opacity
    (\citealt{WyitheBolton11}) and the line-of-sight proximity effect
    (\citealt{Calverley11}).  {\it Middle:} the filled symbols (offset
    from $z=7$ for clarity) display the corresponding volume averaged
    \HI fraction.  The open symbols show the \HI fraction which
    results if self-shielding (see Eq.~\ref{eq:shield}) is not
    included.  Observational constraints at $z\leq 6.1$ are from IGM
    \Lya absorption measurements
    (\citealt{Fan06b,BoltonHaehnelt07b,McGreer11}). {\it Bottom:} the
    number density of absorption systems with $10^{17.2}\leq N_{\rm
      HI}/{\rm cm^{-2}}\leq 10^{20.0}$ per unit redshift.  The filled
    (open) symbols at $z=7$ again show the model predictions including
    (excluding) gas which is self-shielded from the ionising
    background.  The data points with error bars display the number
    density of absorbers per unit redshift measured by
    \citet{Songaila10} at $z<6$, rescaled to match the cosmological
    parameters assumed in this work.}
  \label{fig:obs}
\end{center}
\end{figure}

\begin{figure*}
\centering
\begin{minipage}{180mm}
\begin{center}
 \psfig{figure=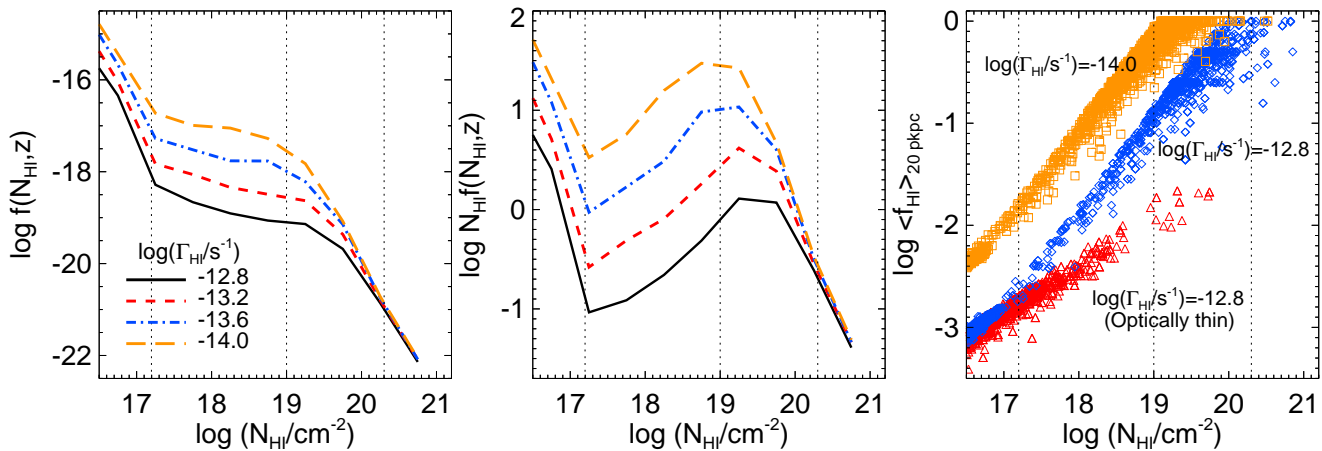,width=1.0\textwidth}
 \vspace{-0.6cm}

  \caption{{\it Left:} the column density distribution function
    predicted at $z=7$ for the four different background ionisation
    rates indicated on the plot.  The vertical dotted lines demarcate
    the column density thresholds by which we define LLSs
    ($10^{17.2}\leq N_{\rm HI}/{\rm cm^{-2}} < 10^{19.0}$), super-LLSs
    ($10^{19.0}\leq N_{\rm HI}/{\rm cm^{-2}} < 10^{20.3}$) and DLAs
    ($N_{\rm HI} \geq 10^{20.3}\rm\,cm^{-2}$).  {\it Middle:} as for
    left panel but now displaying $\log N_{\rm HI}f(N_{\rm HI})$.
    {\it Right:} scatter plots of absorber \HI fractions against \HI
    column density for the largest and smallest of the ionisation
    rates assumed.  The red triangles show the results obtained when
    self-shielding is ignored for $\log(\Gamma_{\rm HI}/\rm
    s^{-1})=-12.8$ only.}
  \label{fig:cddf}

\end{center}
\end{minipage}
\end{figure*}

It is instructive to first make contact with observations by comparing
our simulation results at $z=7$ to constraints from the IGM at $z\leq
6$.  The upper panel in Figure~\ref{fig:obs} displays the four
photo-ionisation rates used in Figure~\ref{fig:slices}.  The filled
symbols in the middle panel display the corresponding volume averaged
\HI fractions predicted by these models.  The lowest \HI fraction is
consistent with only weak evolution from the lower observational limit
at $z=6$, whereas smaller photo-ionisation rates result in a stronger
increase in the \HI fraction toward $z=7$.  Note these \HI fractions
are higher than the values expected if self-shielding were ignored --
this is demonstrated by the open symbols at $z\sim7$, which display
the \HI fraction computed from our simulated sight-lines without the
self-shielding prescription given by Eq.~(\ref{eq:shield})
(i.e. assuming a uniform, optically thin ionising background).

The bottom panel of Figure~\ref{fig:obs} displays the resulting number
density of absorption systems with column\footnote{Column densities
  are estimated by integrating the density field over a scale of 20
  pkpc; this choice is motivated by Eq.~(\ref{eq:Jeans}), which yields
  $L_{\rm ss}\simeq 16$--$40$ pkpc for the range of photo-ionisation
  rates considered here.} densities $10^{17.2}\leq N_{\rm HI}/{\rm
  cm^{-2}} \leq 10^{20.0}$, along with the recent observational
measurements presented by \cite{Songaila10} at $z\leq 6$.  The filled
(open) symbols again display the results including (excluding)
self-shielding.  Lowering the amplitude of the \HI ionising background
significantly increases the incidence of these absorption systems.
This increase is more pronounced in the models incorporating
self-shielding, with a factor of $\sim 25$ increase in the number
density of absorbers per unit redshift from $\log(\Gamma_{\rm HI}/{\rm
  s^{-1}})=-12.8$ to $-14.0$.

The absorber properties are displayed in more detail in
Figure~\ref{fig:cddf}, where the left panel displays the \HI column
density distribution functions (CDDFs) obtained from the simulations.
The CDDF follows the conventional definition:

\begin{equation} f(N_{\rm HI},z) = \frac{\partial^{2}N}{\partial z \partial N_{\rm HI}}\frac{H(z)}{H_{0}(1+z)^{2}}. \end{equation}

\noindent
Including self-shielding increases the number of optically thick
absorbers, flattening the otherwise power-law column density
distribution at $10^{17.2}\leq N_{\rm HI}/{\rm cm^{-2}} \leq
10^{19.2}$.  A similar flattening of the CDDF is inferred from
observations at $z\sim 3.7$ (\citealt{Prochaska10}), and it has also
been noted in many other theoretical models
(\citealt{Katz96b,ZhengMiralda02,Nagamine10,McQuinn11,Altay11}).  The
distributions converge at $N_{\rm HI}\sim 10^{20}\rm\,cm^{-2}$,
indicating the highest column density systems (damped \Lya absorbers
with $N_{\rm HI}\geq 10^{20.3}\rm\,cm^{-2}$) remain largely
insensitive to changes in the background photo-ionisation rate by
virtue of their high gas density, at least for $\Gamma_{\rm HI} \leq
10^{-12.8}\rm\,s^{-1}$.  The middle panel in Figure~\ref{fig:cddf}
displays $N_{\rm HI}f(N_{\rm HI})$, which indicates which optically
thick systems comprise the bulk of the \HI opacity.  When
self-shielding is included, it is absorption systems with $N_{\rm
  HI}\sim 10^{18.5}$--$10^{19.5}\rm\,cm^{-2}$ which dominate.

Finally, the ionisation state of the optically thick absorbers are
displayed in the right panel of Figure~\ref{fig:cddf} (see also figure
5 of \citealt{McQuinn11}).  The blue diamonds and orange squares show
the average \HI fraction against column density for $\log(\Gamma_{\rm
  HI}/\rm s^{-1})=-12.8$ and $-14.0$, respectively.  As the
photo-ionisation rate is lowered, the density threshold for
self-shielding is reduced (since $\Delta_{\rm ss}\propto
\Gamma_{-12}^{2/3}$) and the size and incidence rate of optically
thick absorption systems (i.e. $N_{\rm HI}>\sigma_{\rm
  HI}^{-1}=10^{17.2}\rm\,cm^{-2}$, where $\sigma_{\rm HI}$ is the \HI
photo-ionisation cross-section at the Lyman limit) increases.  The
self-shielded regions first extend into the outer parts of galactic
haloes and then into the filaments defining the cosmic web
(e.g. Figure~\ref{fig:slices}).  For comparison, the red triangles
display the results obtained for $\log(\Gamma_{\rm HI}/\rm
s^{-1})=-12.8$ when ignoring self-shielding.  Absorbers with the same
$N_{\rm HI}$ have larger \HI fractions when self-shielding is
included, implying these must correspond to regions of lower gas
density relative to the optically thin case.  As we shall now
demonstrate, it is this rapid increase in the incidence of optically
thick absorption systems with increasing neutral fraction at the
tail-end of reionisation which is primarily responsible for
strengthening the red \Lya damping wing in our simulations.

\section{Implications for LAE visibility}
\subsection{Photo-ionisation by a central source}

\begin{figure*}
\centering
\begin{minipage}{180mm}
\begin{center}
 \psfig{figure=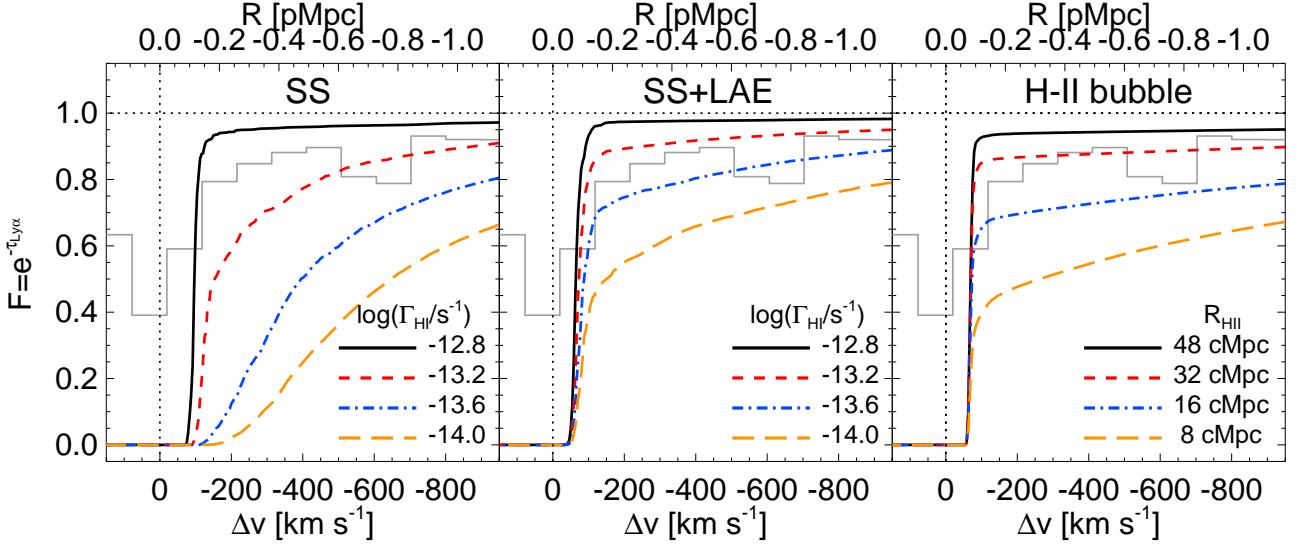,width=1.0\textwidth}
 \vspace{-0.6cm}
 
  \caption{The median \Lya transmission redward of the rest frame
    Ly$\alpha$ transition (at $\Delta v = 0\rm\,km\,s^{-1}$) obtained
    from an ensemble of 300 simulated sight-lines.  Each panel
    displays a different intergalactic Ly-$\alpha$ absorption model.
    The four continuous curves in each panel display the transmission
    profile for different background ionisation rates as indicated.
    The binned grey line displays the observed transmitted fraction
    around the $z=7.085$ quasar ULAS J1120$+$0641 with an estimated
    ${\dot N_{\rm ion}}=1.3\times 10^{57}\rm s^{-1}$.  {\it Left:}
    model which includes self-shielded (SS) gas, but ignores the
    influence of a central ionising source.  {\it Centre:} as for left
    panel, but now each sight-line includes the photo-ionisation rate
    from an LAE with $L_{\rm Ly\alpha}=5\times
    10^{42}\rm\,erg\,s^{-1}$ which emits ionising photons at a rate
    ${\dot N}_{\rm ion}=1.1\times 10^{53}\rm\,s^{-1}$. {\it Right:} a
    model which neglects optically thick systems, and instead models
    the ionisation state of the IGM as an ionised bubble with radius
    $R_{\rm HII}$ embedded in an otherwise fully neutral IGM. The \HII
    bubble sizes are as indicated on the plot.  Note this case is
    shown to emulate results from existing models in the literature
    only.}

  \label{fig:wings}
\end{center}
\end{minipage}
\end{figure*}

We now turn to the main result in this paper; estimating the impact of
the increasing incidence of optically thick absorbers on the
visibility of LAEs.  Before proceeding further, however, we also allow
for the possible effect of ionising radiation from a central source in
our models (i.e. the \Lya emitter under observation).  Ionising
radiation from the LAE itself can reduce the incidence of proximate
self-shielded gas and therefore weaken the \Lya damping wing.
 
We model the possible effect of the LAE as follows.  Firstly, we
compute the emission rate of ionising photons into the IGM, ${\dot
  N}_{\rm ion}\, [\rm s^{-1}]$, by an LAE with observed \Lya
luminosity $L_{\rm Ly\alpha}$ assuming case-B recombination:

\begin{equation} {\dot N}_{\rm ion} = \frac{3}{2}\frac{L_{\rm Ly\alpha}}{h_{\rm P}{\nu_{\alpha}}}\frac{f_{\rm esc}}{(1-f_{\rm esc})f_{\alpha}}. \label{eq:nion} \end{equation}

\noindent
Here the energy of a \Lya photon is $h_{\rm
  P}\nu_{\alpha}=10.2\rm\,eV$, and $f_{\rm esc}$ and $f_{\alpha}$ are
the escape fractions of Lyman continuum and \Lya photons into the IGM.
In this work we follow \cite{Dijkstra07b} and assume $f_{\alpha}=1$,
such that $L_{\rm Ly\alpha} = f_{\alpha}L_{\rm Ly\alpha}^{\rm intr}$
yields the \Lya luminosity uncorrected for dust.  We adopt $L_{\rm
  Ly\alpha}=5\times 10^{42}\rm\,erg\,s^{-1}$ based on the typical \Lya
luminosities measured in recent observations of $z\sim 7$ LAEs
(e.g. \citealt{Pentericci11}).  We assume a Lyman continuum escape
fraction of $f_{\rm esc}=0.2$, which yields ${\dot N}_{\rm
  ion}=1.1\times 10^{53}\rm\,s^{-1}$.  The Lyman continuum escape
fraction is highly uncertain.  A value $\ga 20$ per cent is required
at $z \simeq 6$ if the observed Lyman break galaxy population is to
reproduce ionising background constraints from the \Lya forest
(\citealt{BoltonHaehnelt07b,Shull12,Finkelstein12}), but as we will
discuss in more detail in Section~\ref{sec:cav} the relation between
\Lya emission and the escape of ionising radiation may not be
straightforward.

In the models with a central ionising source we then assume the total
photo-ionisation rate to be the sum of the LAE contribution and the
metagalactic background rate:

\begin{equation} \Gamma_{\rm HI}^{\rm tot}(R) = \sigma_{\rm HI}\frac{{\dot N}_{\rm ion}}{4\pi R^{2}} \frac{\beta}{\beta+3} + \Gamma_{\rm HI}, \label{eq:gamma} \end{equation}

\noindent
where we have assumed a power law spectral energy distribution for the
ionising radiation from the LAE, $L_{\nu}\propto \nu^{-\beta}$, with
$\beta=3$ (cf. \citealt{Leitherer99}).  Note that we have ignored (i)
the finite lifetime of the LAE, effectively assuming the ionising
radiation propagates instantaneously along the line-of-sight (ii) the
intervening \HI opacity between self-shielded systems and (iii)
assumed the LAE luminosity is constant.  All three assumptions will
maximise the impact of ionising radiation from the LAE on the
surrounding neutral hydrogen.  In what follows we therefore consider
this as a limiting case and bracket other possibilities by
additionally computing the \Lya damping wing when the LAE is ignored.
We do not explicitly account for the possible clustering of ionising
sources around the LAE in addition to the range of background
photo-ionisation rates we already consider.  As we shall demonstrate
in the next section, larger background photo-ionisation rates weaken
the red \Lya damping wing by raising the density threshold at which
gas self-shields.

\subsection{The red \Lya damping wing}

Figure~\ref{fig:wings} displays the median IGM \Lya transmission
profile obtained from the 300 simulated sight-lines -- each panel
corresponds to a different IGM absorption model.  Note that when
computing the transmission profiles we have ignored the contribution
of all gas within $20$ pkpc of the centre of the LAE host halo to both
the Lyman continuum and \Lya opacity.  We do not model the complex
radiative transfer within the host galaxy's interstellar medium, and
account for the absorption by this material with the parameters
$f_{\rm esc}$ and $f_{\alpha}$ instead.

The left panel shows the transmission profile predicted by our
simulations without photo-ionisation by the LAE.  A strong damping
wing profile is produced, even for photo-ionisation rates that
correspond to \HI fractions as low as $\sim 3$ per cent (blue
dot-dashed curve).  This is due to the presence of high column density
LLS or damped \Lya absorption systems within $\sim 1$ pMpc of the LAE
host halo (at $\Delta v = 0\rm\,km\,s^{-1}$).  For comparison, the
grey binned line in all panels displays the transmission profile of
the $z=7.085$ quasar ULAS J1120$+$0641 (\citealt{Mortlock11}); a
strong damping wing extends redwards of systemic \Lya despite the
expected quasar proximity effect and the fact that the biased regions
where bright quasars form should reionise earlier than on average.
This implies that if similar ionisation conditions hold around bright
LAEs at $z=7$, an even stronger red damping wing may result.

The central panel instead shows the damping wings for our models
including ionising emission from an LAE.  The source dominates the
total photo-ionisation rate (i.e. produces more than 50 per cent of
the total) within $\sim 0.5 \rm \, pMpc$.  The incidence of proximate,
self-shielded systems within this radius is therefore significantly
reduced, and the \Lya damping wing is consequently weakened.  On the
other hand, recall that the typical halo mass in our simulations is
rather small (see Section~\ref{sec:los}).  We therefore likely
underestimate the abundance of proximate optically thick systems and
this effect may be somewhat less important.  The convergence of the
transmission profiles at $F=0$ (close to $\Delta v =
-50\rm\,km\,s^{-1}$) is due to peculiar velocities arising from
gravitational infall around the haloes.  We have verified that
ignoring peculiar motions produces profiles which converge at $\Delta
v=0\rm\,km\,s^{-1}$ instead.  Note that in the left panel of
Figure~\ref{fig:wings}, where emission from a central LAE is excluded,
this convergence is less prominent.  This is because absorption line
cores from proximate absorbers in addition to the \Lya damping wing
can extend redward of systemic Ly$\alpha$; this occurs more frequently
as the photo-ionisation rate is lowered and the size of the absorbers
increases.

\begin{figure}
\begin{center}
  \includegraphics[width=0.48\textwidth]{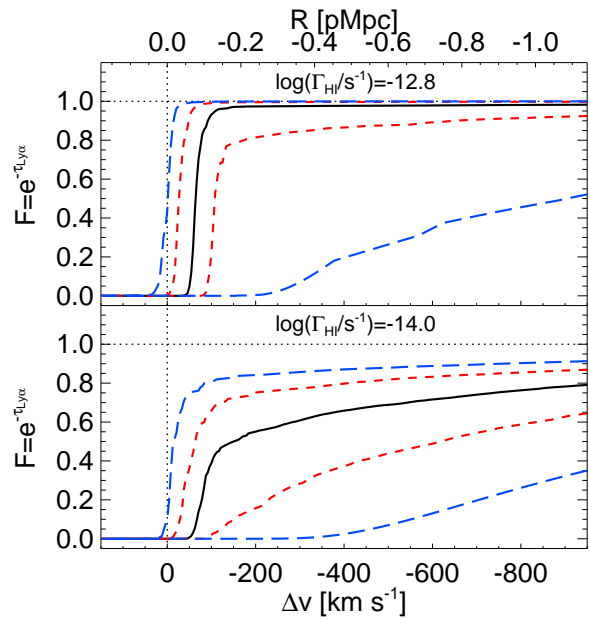}
  \vspace{-0.65cm}
  \caption{The scatter in \Lya transmission along individual
    sight-lines. Each panel assumes the background photo-ionisation
    rates indicated on the diagram and include a central ionising
    source with ${\dot N}_{\rm ion}=1.1\times 10^{53}\rm\,s^{-1}$.
    The short-dashed (red) and long-dashed (blue) curves bound 68 and
    95 per cent of the \Lya transmission around the median (black
    curves).  Note the median profiles are identical to those
    displayed in the central panel of Figure~\ref{fig:wings}.}
  \label{fig:scatter}
\end{center}
\end{figure}

At this point, it is important to emphasise that the transmission
profiles also vary significantly between individual sight-lines; a
strong damping wing will occur whenever there are super-LLS or damped
\Lya absorbers lying close to the host halo.  This can occur even when
the IGM is highly ionised (i.e. for $\log(\Gamma_{\rm
  HI}/\rm\,s^{-1})=-12.8$) assuming a sufficiently overdense region
lies in close proximity to the LAE.  This variation can be seen
clearly in Figure~\ref{fig:scatter}, which shows the 68 and 95 per
cent bounds for the scatter around the median \Lya absorption profile
for $\log(\Gamma_{\rm HI}/\rm s^{-1})=-12.8$ (upper panel) and $-14.0$
(lower panel).  Some \Lya emitters will therefore be completely
obscured by the damping wing while others will remain fully visible.
The stochastic nature of the damping wing absorption from these high
column density systems should therefore also impact on the clustering
of LAEs (e.g. \citealt{Furlanetto06,McQuinn07b}) for even relatively
low \HI fractions of $3$--$10$ per cent, modifying the enhanced
clustering signal expected as the IGM \HI fraction rises.  This will
be an interesting possibility to explore in future work.

In summary, it is clear optically thick absorbers can produce a rapid
decline in the transmission of \Lya emission from high-redshift
galaxies at the tail-end of reionisation. This is quite different to
existing predictions using large-scale numerical and semi-analytical
simulations of reionisation.  In the late stages of reionisation the
presence of self-shielded regions where hydrogen can recombine and
stay neutral is very important
(\citealt{MiraldaEscude00,Ciardi06,GnedinFan06,Choudhury09,Crociani11,AlvarezAbel12}).
Large-scale reionisation simulations do not yet have the dynamic range
to resolve these systems (see e.g. \citealt{TracGnedin11}).  Any
neutral gas is thus typically far away from the central regions of
\HII regions, and the impact of nearby optically thick absorbers on
the \Lya damping wing is ignored.

We illustrate this in the right-hand panel of Figure~\ref{fig:wings},
where we have simulated the \Lya damping wing expected around a LAE
embedded in an \HII bubble with radius $R_{\rm HII}$.  Inside the
bubble ($R\leq R_{\rm HII}$), the IGM is highly ionised with a uniform
\HI fraction of $10^{-4}$, while outside the IGM is fully neutral.
This simple model serves to illustrate the importance of optically
thick absorbers for the red damping wing.  Bubble sizes of $R_{\rm
  HII}\la 30$ cMpc are required to produce a damping wing similar in
strength to the absorption seen when the IGM is only $\sim 3$ per cent
neutral by volume (blue dot-dashed curve in the central panel of
Figure~\ref{fig:wings}) when including optically thick absorbers.  In
comparison, large scale reionisation simulations typically predict
\HII bubble sizes of $\sim 30$--$40$ cMpc for \HI fractions as large
as 30 per cent (e.g. figure 2 in \citealt{Zahn11}).  This elucidates
why existing models indicate the LAE/LBG fraction at $z\sim 7$ may
imply large \HI fractions of 40-90 per cent.  At fixed neutral
fraction, large-scale reionisation simulations predict a damping wing
which is significantly weaker compared to the case where absorption by
optically thick absorbers is included. The damping wing will therefore
only reduce LAE visibility at times when individual \HII bubbles have
not yet grown to the typical size of $\ga 50$ cMpc reached at overlap
in these simulations.

\begin{figure*}
\centering
\begin{minipage}{180mm}
\begin{center}
 \psfig{figure=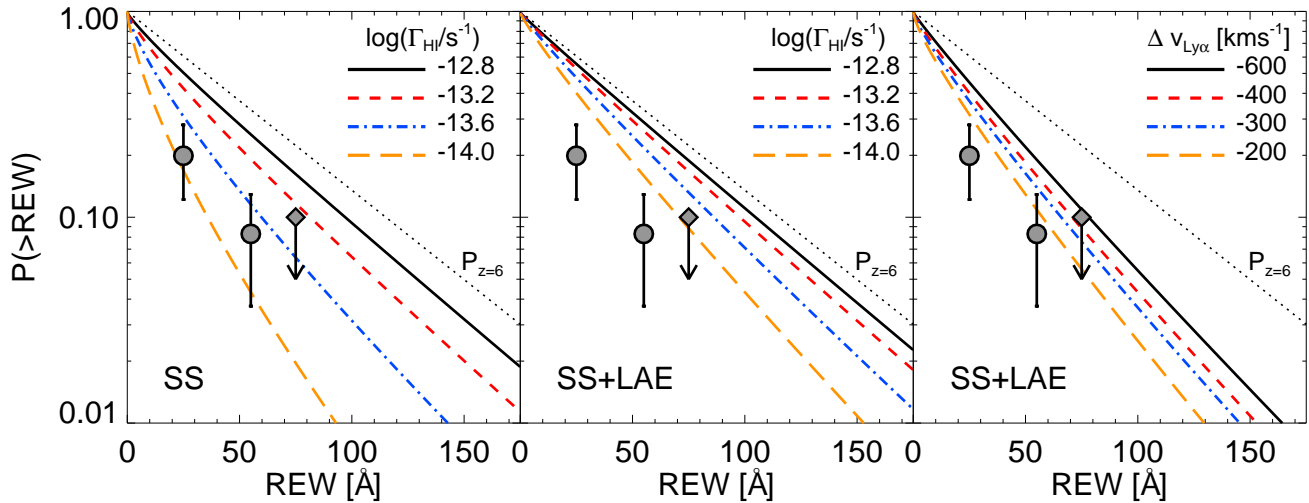,width=1.0\textwidth}

  \vspace{-0.2cm}
  \caption{Predicted cumulative probability distributions (CPDFs) for
    the rest frame \Lya equivalent width (REW) at $z=7$.  The dotted
    curves display the intrinsic $z=6$ CPDF assuming the IGM is fully
    transparent to \Lya photons \citep{Dijkstra11}.  The solid,
    short-dashed, dot-dashed and long-dashed curves are constructed
    from our \Lya transmission models at $z=7$ excluding (left) and
    including (centre) ionising radiation emitted along the line of
    sight from a source with $L_{\rm Ly\alpha}=5\times
    10^{42}\rm\,erg\,s^{-1}$ and ${\dot N}_{\rm ion}=1.1\times
    10^{53}\rm\,s^{-1}$ (see text for details).  These models adopt an
    intrinsic REW distribution identical to that at $z=6$ and differ
    due to evolution in the \Lya transmission only.  A Gaussian \Lya
    profile with a standard deviation of $88\rm\,kms^{-1}$ and an
    offset of $-400\rm\,km\,s^{-1}$ redward of rest frame \Lya has
    been assumed in both cases.  The right hand panel instead displays
    the $\log(\Gamma_{\rm HI}/\rm s^{-1})=-14.0$ case from the central
    panel (with $\langle f_{\rm HI} \rangle_{\rm V}\sim 0.1$) assuming
    four different \Lya emission line offsets. The models are compared
    to the composite data set compiled by \citet{Ono12} (filled
    circles) which utilises observations from
    \citet{Fontana10,Vanzella11,Pentericci11} and \citet{Schenker12}.
    The upper limit at $75\rm\,\AA$ is based on the earlier
    non-detection by \citet{Stark10} at $z\sim 7$.}

  \label{fig:REW}
  
\end{center}
\end{minipage}
\end{figure*}

\subsection{The REW distribution}

We now compare our models directly to recent measurements of the
LBG/LAE fraction
(\citealt{Fontana10,Vanzella11,Pentericci11,Ono12,Schenker12})
following the approach of \cite{Dijkstra11}, which we recapitulate
here.  Our first requirement is to compute the fraction of the \Lya
emission line which is transmitted, ${\mathscr T}$, where

\begin{equation} {\mathscr T} = \frac{\int_{\nu_{\rm min}}^{\nu_{\rm max}} d\nu J(\nu)e^{-\tau_{\rm Ly\alpha}(\nu)}}{\int_{\nu_{\rm min}}^{\nu_{\rm max}}d\nu J(\nu)}.\end{equation}

\noindent
Here $J(\nu)$ gives the shape of the \Lya line profile, $e^{-\tau_{\rm
    Ly\alpha}(\nu)}$ is the IGM transmission
(e.g. Figure~\ref{fig:wings}) and $\nu_{\rm min}$ and $\nu_{\rm max}$
are chosen to correspond to the full extent of our simulated
sight-lines.  To compute ${\mathscr T}$ for each of our simulated
sight-lines we must assume an intrinsic \Lya profile.  We adopt a
Gaussian line profile with a standard deviation of
$88\rm\,km\,s^{-1}$, corresponding to the circular velocity of a dark
matter halo of mass $10^{10}M_{\odot}$ at $z=7$ (eq. 25 in
\citealt{BarkanaLoeb01}). We further assume that the line centre is
shifted redward of \Lya by $400\rm\,km\,s^{-1}$; this offset is
typical of \Lya emission lines associated with LBGs at $z\simeq 2-3$
(\citealt{Steidel10}), and is chosen to mimic the effect of resonant
scattering by the (outflowing) gas on the \Lya line visibility (but
see our discussion in the next section).

We next construct a probability distribution $dP(<\log{\mathscr
  T})/{d\log\mathscr T}$ for each IGM absorption model using our
simulations.  For \Lya redshifted relative to systemic we assume the
observed rest frame equivalent width (REW) at $z=6$, when the hydrogen
in Universe is already highly ionised, to be equal to the emitted REW
(i.e. ${\mathscr T}_{z=6}=1$). If we further assume that there is no
intrinsic evolution in the LAE population from $z=6$ to $z=7$, the
cumulative probability distribution (CPDF) that a galaxy has an
observed REW and an IGM transmission $\mathscr T$ at $z=7$ is

\begin{equation} P(>{\rm REW}) = \int_{0}^{1} d{\mathscr T} \frac{dP(<\log{\mathscr T} )}{d\log{\mathscr T}} \frac{P_{z=6}(>{\rm REW}/{\mathscr T})}{{\mathscr T}\ln 10}. \end{equation}

\noindent
Here ${\rm REW}_{\rm intr} ={\rm REW}/{\mathscr T}$, $P_{z=6}(>{\rm
  REW}) = \exp({\rm -REW/REW_{\rm c}})$ and $\rm REW_{\rm
  c}=50\rm\,\AA$ (see \citealt{Dijkstra11} for details).

Figure~\ref{fig:REW} displays the resulting CPDFs for the IGM models
with self-shielding only (left panel) and including ionisation by a
central LAE (central panel).  Note again these two cases have been
chosen to bracket the uncertain impact of Lyman continuum emission
from the central LAE on the surrounding IGM (see also the discussion
in Section~\ref{sec:cav}).  The CPDFs are compared to a recent data
compilation presented by \cite{Ono12}, as well as an upper limit at
$75\rm\,\AA$ from \cite{Stark10}.  All curves in the central panel lie
above the quoted observational uncertainties, implying a lower
line-of-sight escape fraction ($f_{\rm esc}<0.2$) is required to match
the data if the \HI fraction remains fixed.  As expected, however, our
simulations do predict a rapid evolution in the REW distribution from
$z=6$, even for modest neutral fractions of $3$ per cent (blue
dot-dashed curve).  In the right panel, the CPDF for the
$\log(\Gamma_{\rm HI}/\rm s^{-1})=-14.0$ model in the central panel
(orange long-dashed curve with $\langle f_{\rm HI} \rangle_{\rm V}\sim
0.1$) is shown assuming four different offsets for the \Lya emission
line relative to systemic.  Shifting the line blueward (redward) of
the fiducial $-400\rm\,km\,s^{-1}$ decreases (increases) ${\mathscr
  T}$, producing a stronger (weaker) evolution in the REW
distribution.

\subsection{Uncertainties} \label{sec:cav}

From our discussion thus far, it should be clear that modelling the
IGM transmission for LAEs at the tail-end of reionisation is uncertain
mainly because of the difficulties with self-consistently modelling
self-shielded regions.  However, uncertainties associated with the
escape of ionising radiation from the (possibly clustered) \Lya
emitters and the redshift relative to systemic of the \Lya emission
also play a role.

When including an LAE in our models we have assumed $f_{\rm esc}=0.2$.
It is, however, not obvious that high redshift galaxies emit ionising
and \Lya radiation at the same time and/or in the same direction.  It
is certainly plausible that no ionising radiation escapes in the
direction of the observer (e.g. \citealt{Gnedin08}), in which case
even our IGM model excluding a central source (i.e. the left panel in
Figure~\ref{fig:REW}) may actually be an overestimate of the IGM
transmission ${\mathscr T}$; recall we have ignored all neutral gas
within $20\rm\,pkpc$ of the halo centre and instead assumed the
appropriate escape fractions for ionising and \Lya photons.  Note
further that the escape of ionising radiation and the emission
redshift relative to systemic may also be unfavourably coupled.  A
large redshift for the emission line may need a large column density
of neutral gas within the host halo of the LAE.  This is required in
order to resonantly scatter the \Lya photons locally far into the red
wing of the line (e.g. \citealt{Barnes11}).  The redshifts of
$400\rm\, km\,s^{-1}$ or more for the observed \Lya emission in $z=3$
LBGs (\citealt{Steidel10}) may therefore be incompatible with a
simultaneous escape of ionising radiation along the line-of-sight to
the observer.

The detailed intrinsic line \Lya shape is also rather uncertain -- in
this work we have assumed a simple Gaussian line profile with standard
deviation $88\rm\,km\,s^{-1}$.  In practice, however, radiative
transfer models predict a characteristic double peaked emission line
(e.g. \citealt{Verhamme06}) with intergalactic \Lya absorption
removing the blue peak at high redshift.  In their recent analysis of
the REW distribution predicted by large-scale reionisation
simulations, \cite{Jensen12} adopt a doubled peaked Gaussian minus a
Gaussian line based on the radiative transfer simulations presented by
\cite{Laursen11}.  They note that using this line profile weakens the
impact of the \HI fraction on the REW distribution compared to a
simple Gaussian centred at the systemic redshift.  Note, however,
since we adopt a profile which is shifted redward of systemic, our
assumption that the entire line is visible in the absence of a red
\Lya damping wing is reasonable.  Unless the line width becomes
comparable to the line offset relative to systemic, this will not have
a significant effect on the resulting REW distribution.

Lastly, the small simulation box size of $10h^{-1}\rm\,cMpc$ and the
omission of self-consistent modelling for the ionising sources remains
the most significant concern in this analysis.  We do not correctly
model large-scale variations in the IGM ionisation state prior to
overlap.  On the other hand, we argue that the proper modelling of
dense, self-shielded regions are an equally important part of this
problem.  A fully self-consistent model will require combining both of
these aspects, either by resolving all relevant physical scales or
adopting sub-grid models which are informed by high resolution
simulations of the IGM density field.  Furthermore, any modelling
should be calibrated to reproduce the observed transmission redwards
of \Lya in bright high-redshift quasars which appears to evolve
rapidly between $z=6$--$7$; the simulations presented here have been
used to estimate a neutral fraction of $>10$ per cent around the
$z=7.085$ quasar ULAS J1120$+$0641
(e.g. \citealt{Mortlock11,Bolton11}).  Note, however, this is based on
a small number of bright quasars at $z\sim6$ and only one at $z>7$.
Future observations will help determine whether ULAS J1120$+$0641 is
typical.

\section{Conclusions}

We have used a hydrodynamical simulation to model the \Lya opacity of
the intervening IGM during the final stages of reionisation.  As the
photo-ionisation rate drops the opacity redward of rest-frame \Lya is
expected to rise rapidly due to the increasing incidence of optically
thick absorption systems.  Our results indicate that the bulk of the
\HI opacity will arise from optically thick systems with column
densities $N_{\rm HI}\sim 10^{18.5}$--$10^{19.5}\rm\,cm^{-2}$.  When
including these absorption systems in the simulations, only a moderate
rise in the volume averaged neutral fraction is required to
significantly reduce the transmission redward of rest-frame
Ly$\alpha$.

This result has an important implication for the interpretation of the
recently observed decline in the \Lya emission from high-redshift
galaxies at $z\sim 6$--$7$
(e.g. \citealt{Stark10,Pentericci11,Hayes11,Ono12,Schenker12,CurtisLake12}).
We find that these observations do not require a large neutral
fraction ($\sim 40$--$90$ per cent) in the intervening IGM at $z=7$ as
previously suggested.  Instead, if the rapid decline in the LAE/LBG
fraction is further corroborated, it may instead be indicative of the
rapid decrease of the mean free path of ionising photons expected at
the tail-end of reionisation.  Furthermore, as we find the \Lya
emission from high-redshift galaxies will be suppressed for
volume-averaged neutral fractions of only $3$--$10$ per cent, the
patchiness of reionisation may produce a more modest impact on the
clustering properties of observable high-redshift LAEs than previously
predicted.

Our findings may be particularly relevant for future surveys which
plan to use the transmission of \Lya emission and the clustering
properties of LAEs to probe deep into the epoch of reionisation.
Detecting these LAEs at $z>7$ may well become difficult, even when the
IGM is only ten per cent neutral.  It may also mean that
spectroscopic confirmation of high-redshift candidate galaxies
identified with the drop-out technique at $z>7$ may become
problematic, at least until emission lines other than \Lya can be
observed with the James Webb Space Telescope.  Finally, extending
theoretical predictions beyond the approximate approach adopted in
this work is vital.  This will ultimately require incorporating
small-scale absorption systems into simulations which also model the
patchy nature of reionisation on large scales.  This remains a
considerable but fundamental computational challenge for the current
generation of reionisation simulations.

\section*{Acknowledgments}

The hydrodynamical simulation used in this work was performed using
the Darwin Supercomputer of the University of Cambridge High
Performance Computing Service (http://www.hpc.cam.ac.uk/), provided by
Dell Inc. using Strategic Research Infrastructure Funding from the
Higher Education Funding Council for England.  We thank Volker
Springel for making GADGET-3 available, and Mark Dijkstra, Zoltan
Haiman and Matt McQuinn for helpful comments on the draft manuscript.
Figure~\ref{fig:slices} utilises the cube helix colour scheme
introduced by \cite{Green11}.  JSB acknowledges the support of an ARC
postdoctoral fellowship (DP0984947), and thanks Pratika Dayal and
Stuart Wyithe for valuable conversations.

%\bibliographystyle{apj}
%\bibliography{/Users/jamesbolton/PAPERS/references}

\end{document}